\documentclass[12pt]{iopart}
\usepackage{epsfig}
\usepackage{iopams}
\begin{document}

\title[Exciting the QGP with a relativistic jet]
{Exciting the quark-gluon plasma with a relativistic jet}

\author{Massimo Mannarelli and Cristina Manuel}

\address{Instituto de Ciencias del Espacio (IEEC/CSIC),\\
Campus U.A.B., Fac. de Ci\`encies, Torre C5 E-08193 Bellaterra (Barcelona), Spain}

\begin{abstract}
We discuss the properties of a system composed by a static plasma traversed by a jet of particles.
Assuming that both the jet and the plasma can be described using a hydrodynamical approach, and
in the conformal limit, we find that unstable modes arise when the velocity of the jet is larger than
the speed of the sound of the plasma and only modes with momenta smaller than a certain values are
unstable. Moreover, for ultrarelativistic velocities of the jet the most unstable modes correspond to
relative angles between the velocity of the jet and momentum of the collective mode $\sim \pi/4$.
Our results suggest an alternative mechanism for the description of the jet quenching phenomenon, where
the jet crossing the plasma loses energy exciting colored unstable modes. In LHC this effect should be
seen with an enhanced production of hadrons for some specific values of their momenta and in certain
directions of momenta space.
\end{abstract}


It has been suggested that a high $p_T$ jet crossing the medium produced
 after a relativistic heavy ion collision,
and travelling at a velocity higher than the speed of sound should  form shock waves
with a Mach cone structure \cite{Casalderrey-Solana:2004qm}. 
Such shock waves should be detectable in the
low $p_T$ parton distributions at angles $\pi \pm 1.2 $ with respect to the direction of the trigger particle.
A preliminary analysis of the azimuthal dihadron correlation performed by the PHENIX Collaboration
\cite{Adler:2005ee} seems to suggest  the formation of such a conical flow.

We propose a novel possible collective
process to describe the jet quenching phenomenon. In our approach 
a neutral beam of colored particles crossing
an equilibrated quark-gluon plasma induces plasma instabilities
\cite{Mannarelli:2007gi}. Such instabilities represent  a very efficient mechanism for converting the  energy and momenta stored in the total system
(composed by the plasma and the jet)  into
(growing) energy and momenta of gauge fields, which are
initially absent.  
To the best of our knowledge, only Ref.~\cite{Pavlenko:1991ih} considers the
possibility of the appearance of filamentation instabilities
produced by hard jets in heavy-ion collisions.  

We have studied this phenomenon using the chromohydrodynamical approach developed in \cite{Manuel:2006hg}, assuming the conformal limit for the
plasma. Since we are describing the system employing ideal fluid-like equations,  our results are valid at  time scales shorter than the average time for collisions.  A similar analysis using kinetic theory, and reaching to similar results, will soon be reported.

We have studied the dispersion laws of the gauge collective modes and their
dependence on the velocity of the jet $v$, the magnitude of the momentum
of the collective mode ${\bf k}$, the angle $\theta$ between these quantities,
and of the plasma frequencies of both the plasma $\omega_p$ and the jet $\omega_{\rm jet}$.
We find that there is always one unstable  mode if the velocity of the jet is larger than the speed of sound
$c_s = 1/\sqrt{3}$,
and if the momentum of the collective mode is in modulus smaller than a threshold value.
 Quite interestingly we find that  the unstable modes with momentum parallel to the velocity of the jet
 is the dominant one for velocity of the jet $v \lesssim 0.8$.
For larger values of the jet velocity only the modes with angles larger than $\sim \pi/8$
are significant and the dominant unstable modes correspond to  angles $\sim \pi/4$ (see Figure 1).

Our numerical results imply that both in RHIC and in the LHC these instabilities develop very fast,
faster in the case of the LHC as there one assumes that $\omega_p$ will attain larger values.
Further, the soft gauge fields will eventually decay into soft hadrons, and may affect the hydrodynamical
simulations of shock waves mentioned in reference \cite{Casalderrey-Solana:2004qm}.

\begin{figure}[!th]
\label{plots}
\includegraphics[width=3in,angle=-0]{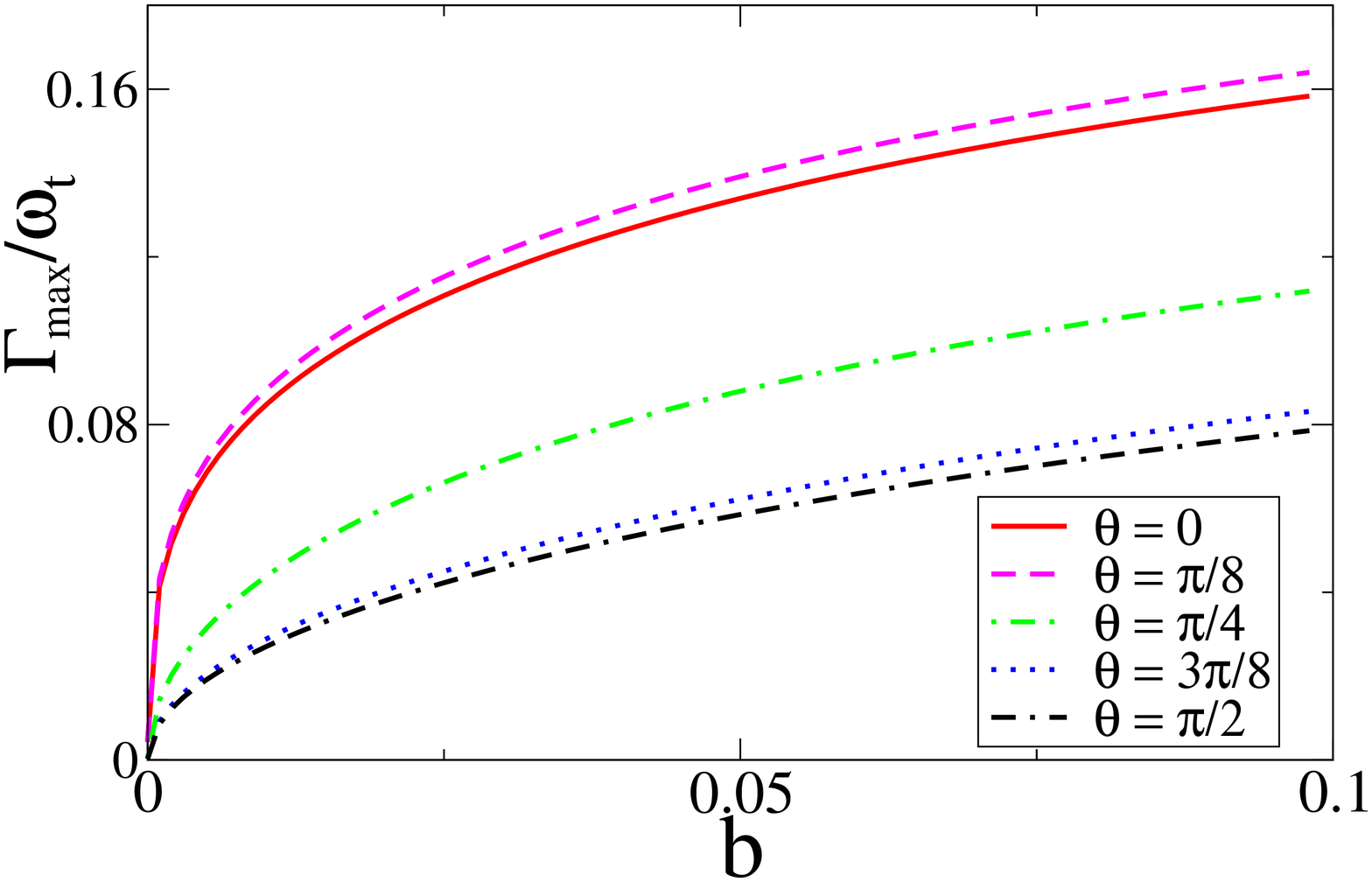}
\includegraphics[width=3in,angle=-0]{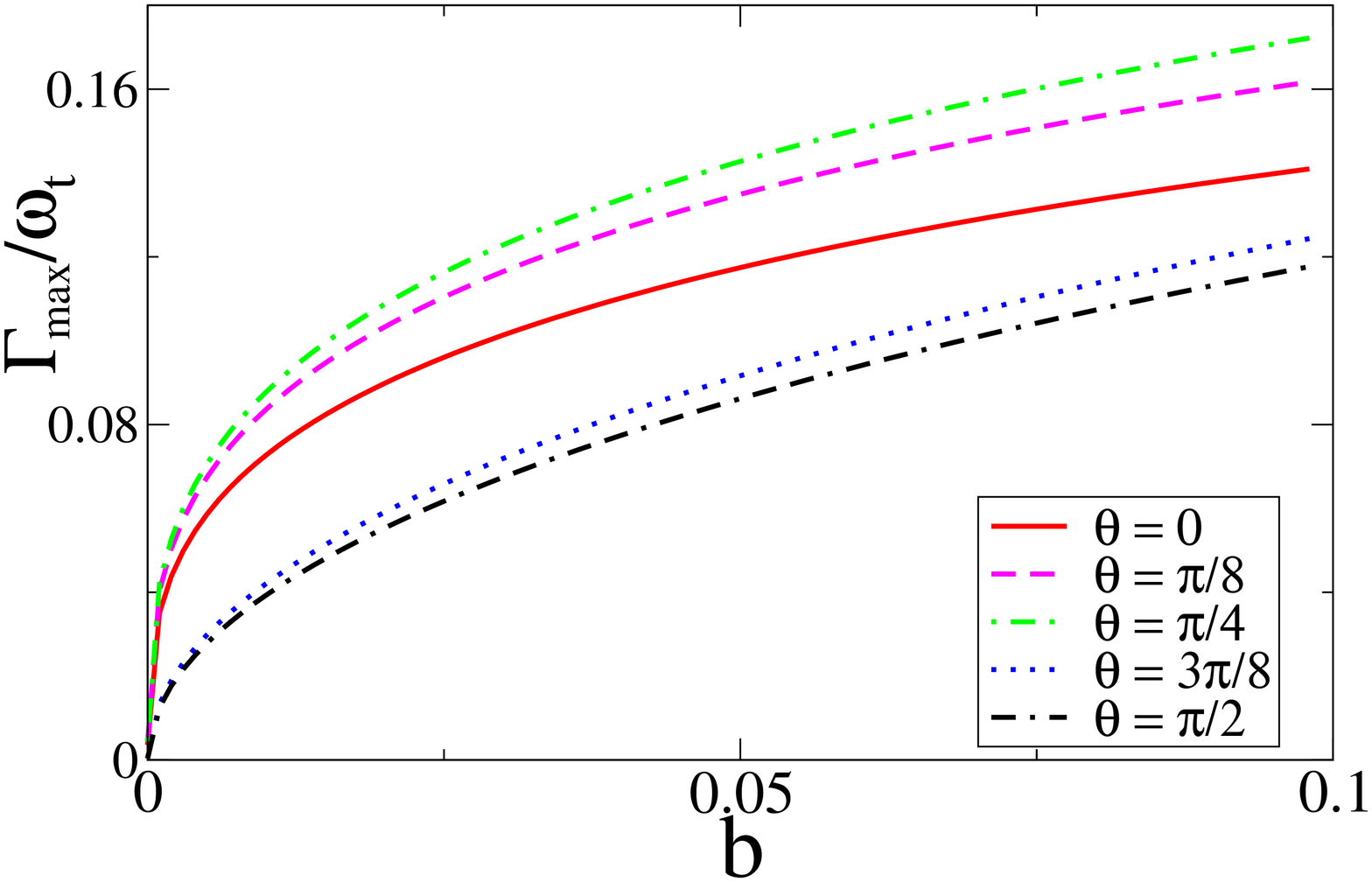}
\caption{ Largest value of the imaginary part of the dispersion law for the unstable mode as a function
of $b= \omega^2_{\rm jet}/\omega_p^2$ for two different values of  the velocity of the jet $v$
 and five different angles between $\bf k$ and $\bf v$. The left/right panels correspond to $v=0.8/0.9$,
 respectively.}
\end{figure}

\ack
 This work has been supported by the Ministerio de Educaci\'on
y Ciencia (MEC) under grant AYA 2005-08013-C03-02.

\end{document}